\documentclass{article}
\usepackage{spconf,amsmath,graphicx}

\usepackage{enumitem}
\setlist{nosep, leftmargin=14pt}

\usepackage{mwe} 
\usepackage{xcolor}
\usepackage{pifont}
\usepackage{cite}
\usepackage{multirow}
\usepackage{amsmath,amssymb,amsfonts}
\usepackage{algorithmic}
\usepackage{graphicx}
\usepackage{textcomp}
\usepackage{verbatim}

\newcommand{\pub}[1]{{\color{gray}{\tiny{[{#1}]}}}}
\newcommand{\ie}[1]{\textit{i.e.,}}
\newcommand{\xhdr}[1]{\vspace{1.7mm}\noindent{{\bf #1.}}}

\title{FDNet: Frequency Domain Denoising Network For Cell Segmentation in Astrocytes Derived From Induced Pluripotent Stem Cells}
\namea{Haoran Li$^{1}$, Jiahua Shi$^{2}$\thanks{Corresponding author, jiahua.shi@uq.edu.au}, Huaming Chen$^{3}$, Bo Du$^{4}$, Simon Maksour$^{6}$, Gabrielle Phillips$^{5}$,}
\nameb{Mirella Dottori$^{5}$, Jun Shen$^{1}$}
\address{
$^{1}$School of Computing and Information Technology, University of Wollongong, Australia\\
$^{2}$Centre for Nutrition and Food Sciences, University of Queensland, Australia\\
$^{3}$School of Electrical and Information Engineering, University of Sydney, Australia\\
$^{4}$Department of Business Strategy and Innovation, Griffith University, Australia\\
$^{5}$School of Medical, Indigenous and Health Sciences, University of Wollongong, Australia\\
$^{6}$School of Chemistry and Molecular Bioscience, University of Wollongong, Australia}
\begin{document}
%
\maketitle
\begin{abstract}
Artificially generated induced pluripotent stem cells (iPSCs) from somatic cells play an important role for disease modeling and drug screening of neurodegenerative diseases. Astrocytes differentiated from iPSCs are important targets to investigate neuronal metabolism. The astrocyte differentiation progress can be monitored through the variations of morphology observed from microscopy images at different differentiation stages, then determined by molecular biology techniques upon maturation. However, the astrocytes usually ``perfectly'' blend into the background and some of them are covered by interference information (\ie, dead cells, media sediments, and cell debris), which makes astrocytes difficult to observe. Due to the lack of annotated datasets, the existing state-of-the-art deep learning approaches cannot be used to address this issue. In this paper, we introduce a new task named astrocyte segmentation with a novel dataset, called IAI704, which contains 704 images and their corresponding pixel-level annotation masks. Moreover, a novel frequency domain denoising network, named FDNet, is proposed for astrocyte segmentation. In detail, our FDNet consists of a contextual information fusion module (CIF), an attention block (AB), and a Fourier transform block (FTB). CIF and AB fuse multi-scale feature embeddings to localize the astrocytes. FTB transforms feature embeddings into the frequency domain and conducts a high-pass filter to eliminate interference information. Experimental results demonstrate the superiority of our proposed FDNet over the state-of-the-art substitutes in astrocyte segmentation, shedding insights for iPSC differentiation progress prediction.
\end{abstract}
\begin{keywords}
neurodegenerative diseases, Induced pluripotent stem cell, deep learning, image segmentation
\end{keywords}
\section{Introduction}
\label{sec1}

Neurodegenerative diseases, including Alzheimer Disease (AD), Parkinson Disease (PD), Huntington Disease (HD), and Amyotrophic Lateral Sclerosis (ALS), are the disorders characterized by the progressive loss, degeneration, or death of specific neuronal cells in the brain or peripheral organs, as well as the deposition of aggregated proteins~\cite{gitler2017neurodegenerative}. These diseases lead to progressive and fatal decline in cognitive, motor, social behavior, psychiatric deficits, and ultimately impairing an individual's ability to perform daily activities~\cite{dugger2017pathology}.  Neurodegenerative diseases are a major threat to human health with growing public concern worldwide due to their impacts on individuals, families, and healthcare systems~\cite{gitler2017neurodegenerative}. While there are no available treatments to cure neurodegenerative diseases, concerted research efforts are undertaken aiming to understand the underlying mechanisms, identify biomarkers for early diagnosis, and develop targeted therapies to slow or halt disease progression~\cite{sun2021gene}. 

Stem cells have the regenerative potential and the ability to differentiate into various specialized cell types, making it possible to replace damaged or lost neurons, promote neuronal repair and regeneration~\cite{okano2022ipsc}. Induced pluripotent stem cells (iPSCs) are artificially generated from somatic cells with great potentials in modeling diseases and drug screening. The feature of \textit{in vitro} cultivation of iPSC makes it an unlimited source of proliferating cells, overcoming the constraints of confined donor cell availability~\cite{ebert2012induced}. Numerous studies have been reported to utilize iPSCs to model diseases, study disease mechanisms, and screen for potential therapeutics~\cite{okano2022ipsc}.

Technological progression has facilitated the adoption of stem cell models and fulfilled the requirement for real-time, label-free, non-perturbing analytical methods to accurately capture the iPSC differentiation process. Incucyte® is a widely used live-cell imaging system that captures phase contrast images to monitor the morphology change along iPSC differentiation, without ever having to displace cells or disrupt their surroundings. Incorporating artificial intelligence (AI) with the analysis of captured images at various iPSC differentiation stages, it can facilitate a powerful platform for researchers to make data-driven decisions and critical insights into disease processes. However, each phase contrast image captures the whole field of designated area including live cells and interference such as background, dead cells, media sediments and cell debris. A robust and accurate cell segmentation solution is needed to achieve precise restitution of live cells and separate them from interference without sacrificing the boundary details.

To fill the gap aforementioned, we introduce a new task in this paper, named astrocyte segmentation in Incucyte images (termed astrocyte segmentation), which aims to detect astrocytes at different stages of differentiation from Incucyte images. We build the \textbf{\underline{I}}ncucyte \textbf{\underline{A}}strocyte \textbf{\underline{I}}mages 704 (termed as IAI704) dataset, a dataset of Incucyte images at different differentiation stages. Each image contains astrocytes with diverse morphologies along with their corresponding segmentation masks.

Based on this dataset, we introduce a high-throughput \textbf{\underline{F}}requency domain \textbf{\underline{D}}enoising \textbf{\underline{Net}}work (termed as FDNet), which can accurately segment astrocytes differentiated from Huntington's Disease iPSC model. Astrocyte is a type of glial cells that maintains, supports and regulates the functions of neurons, which has attracted increasing interest of neuroscientists to investigate its metabolism in neurodegerative diseases. Our network presents a simple, fast and precise approach for the segmentation of astrocytes from Incucyte images and can be promisingly applied to other differentiated cell types with dataset training. 
\vspace{-0.8cm}
\section{Incucyte Astrocyte Images 704 dataset}\label{sec2}

\subsection{Image Collection}\label{sec2.1}
The Huntington's Disease iPSC lines are provided by Cedars Sinai Medical Center (California, USA) and Professor Lesley Jones' team at Cardiff University, UK. Stem cell cultivation and astrocyte differentiation followed the procedures as previously described~\cite{ng2022optimized}. All iPSC experimental protocols were approved by the University of Wollongong (UOW) Human Research Ethics Committee (Ethics Number: 2020/450). Briefly, HD iPSCs were differentiated into neural progenitor cells and further differentiated into astrocytes driven by a lentiviral vector encoding the transcription factors SOX9 and NFIB. The images of differentiated cells were taken once a day with Incucyte® S3 (Sartorius, USA) to record the process of astrocyte morphology change along the maturation journey within 21 days. The differentiation of mature astrocytes were validated by quantitative polymerase chain reaction and immunocytochemistry (data not shown here). 
\subsection{Dataset statistics}\label{sec2.2}
\begin{figure}[!t]
\centerline{\includegraphics[width=0.5\textwidth]{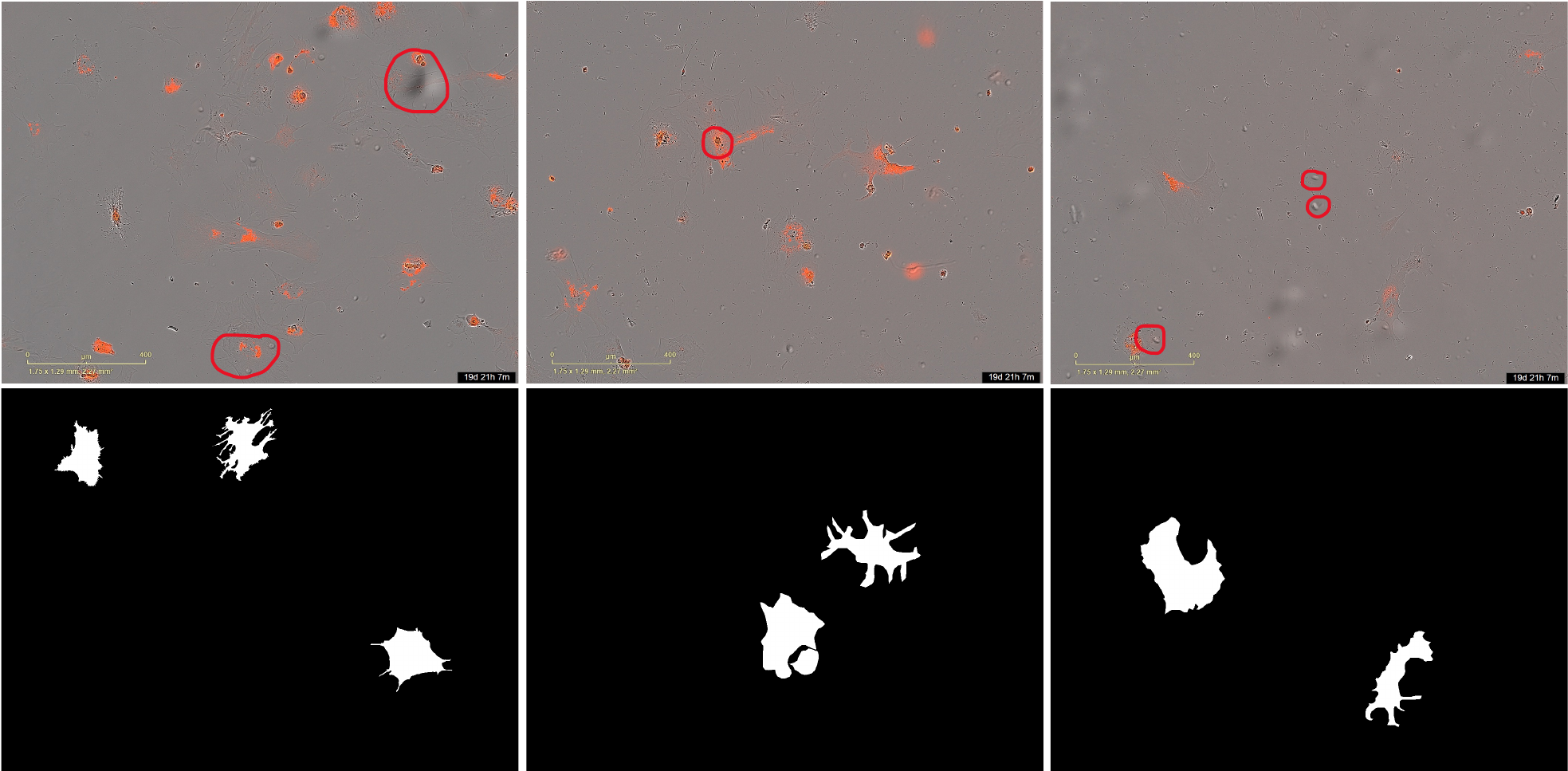}}\vspace{-0.2cm}
\caption{\textbf{Examples of our IAI704.} From top to bottom are original Incucyte images and manually annotated labels. Some examples of interference (\ie, dead cells, media sediments and cell debris) are marked by red circles.}\vspace{-0.2cm}
\label{dataset}
\end{figure}

We now provide some statistics of the proposed dataset. Some examples of the proposed IAI704 are shown in Fig~\ref{dataset}.
\begin{itemize}
    \item[$\bullet$] \textit{Resolution.} Each image collected in IAI704 remains a high-resolution of $1408 \times 1040$ which could provide more details on morphology of the astrocyte. 
    \item[$\bullet$] \textit{Manual Annotation.} Following COCO~\cite{lin2014microsoft}, we annotate all astrocytes from each Incucyte image at an instance-level. Noted that astrocytes with interference information covered are not included in the ground-truth mask.
    \item[$\bullet$] \textit{Dataset splits.} All images from our IAI704 are collected from a 24 well plate. We take the images collected from the first three rows as the train-set and images from the last row as the test-set, which results in 528 images for training and 176 images for testing.
\end{itemize}

\section{Methodology}\label{sec3}

\subsection{Problem Statement}\label{sec3.1}
Given an input dataset $\mathcal{X} = \{(x_i, y_i)\}^{n}_{i=1}$, where $x_i$ denotes the input Incucyte image and $y_i$ denotes the ground-truth segmentation mask. The goal of astrocyte segmentation is to train a segmentation network with $\mathcal{X}$ which could predict the segmentation masks $Y=\{y^{t}_j\}^{n}_{j=1}$ with a new set of Incucyte images $\mathcal{X}^t = \{(x^{t}_j)\}^{n}_{j=1}$ as input.

\begin{figure}[!t]
\centerline{\includegraphics[width=0.5\textwidth]{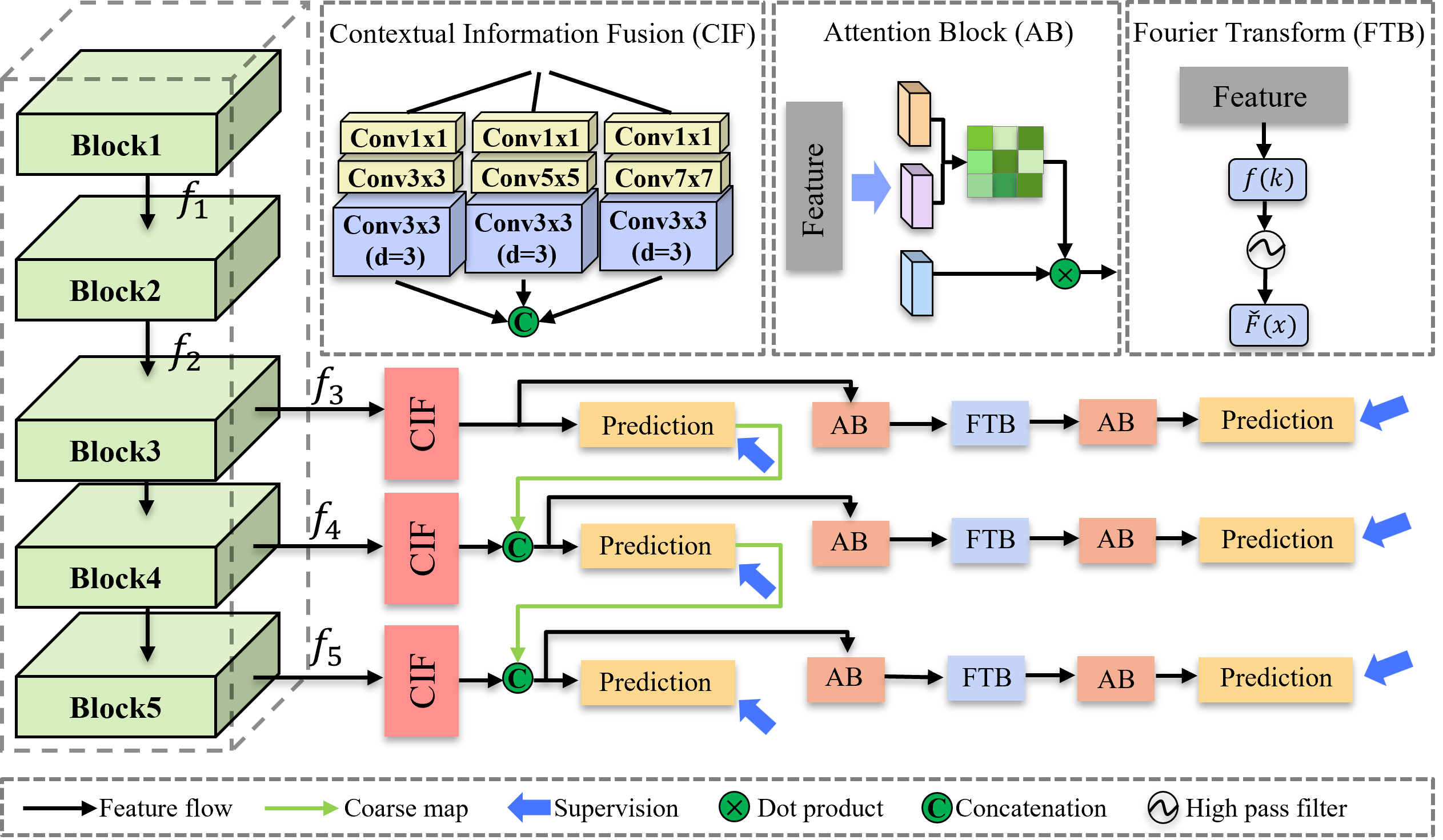}}\vspace{-2.5mm}
\caption{\textbf{Architecture of our proposed FDNet.} Our model contains a backbone, a Contextual Information Fusion module \textit{CIF}, an Attention Block \textit{AB} and a Fourier Transform Block \textit{FTB}.}\vspace{-0.3cm}

\label{framework}
\end{figure}

\subsection{Frequency domain denoising Network}\label{sec3.2}
As shown in Fig.~\ref{framework}, we propose a frequency domain denoising network for astrocyte segmentation in Incucyte images. Our model consists of four key components: backbone, Contextual Information Fusion module (CIF), Attention Block (AB), and Fourier Transform Block (FTB). Specifically, we use ResNet50 as the backbone to extract multi-scale features. Inspired by~\cite{fan2021concealed}, which shows that using multiple atrous convolutions with different rates can expand the receptive field for rich contextual information. Hence we use three parallel atrous convolutions with rates set to 3, 5, and 7 in the proposed CIF module. As mentioned in Sec~\ref{sec1}, due to the similarity between the objects and their surroundings, it is extremely difficult to localize the astrocytes in the Incucyte images. Therefore, after each CIF module, we add a prediction layer to get the coarse result map $y^c_k(k \in {3, 4, 5})$ and concatenate the coarse map with the output feature of the CIF module to get the refined feature $\hat{f}_k (k \in {3, 4, 5})$. AB takes $\hat{f}_k$ as input and applies channel-wise attention for accurate localization. The FTB takes the output of AB and removes the interference through a frequency-level high-pass filter. The output feature map from FTB is then re-sent to AB to get the final feature embedding for prediction.

\subsubsection{Attention Block}\label{sec3.2.1}
Given the input refined feature $\hat{f}_k \in \mathbb{R}^{C \times H \times W}$, where $C$, $H$, and $W$ denote the channel number, height, and width, respectively, we first send it to an average pooling layer and reshape it to get the key $K$, query $Q$ and value $V$. Then a matrix-level multiplication is performed between the query and key to get the channel attention map $M \in \mathbb{R}^{C \times C}$:
\vspace{-2mm}
\begin{equation}
    m_{ij} = \frac{exp(K_{i:} \cdot Q_{j:}^{T})}{\sum^{C}_{j=1}exp(K_{i:} \cdot Q_{j:}^{T})},
\vspace{-2mm}
\end{equation}
where $Q^T$ denotes the transpose of matrix $Q$, $K_{i:}$ and $Q_{j:}^{T}$ denotes the $i^{th}$ row of the matrix $Q$ and the $j^{th}$ row of the matrix $Q^{T}$, and $m_{ij}$ denotes the impact of the $j^{th}$ channel on the $i^{th}$ channel, respectively. Then the output feature map $\widetilde{f}_{k}$ of AB could be calculated through $M$ and $V$:
\vspace{-2mm}
\begin{equation}
    \widetilde{f}_{k} = M \cdot V + \hat{f}_k.
\vspace{-2mm}
\end{equation}

\subsubsection{Fourier Transform Block}\label{sec3.2.2}
As mentioned in Sec~\ref{sec1}, the cells obscured by interference in the Incucyte image are interference information that need to be eliminated. Hence we propose the Fourier Transform Block to remove those cells before getting the final prediction mask. Specifically, our FTB module takes the output feature map $\widetilde{f}_{k}$ of AB as input and uses fast Fourier transform to get the frequency domain map of $\widetilde{f}_{k}$:
\vspace{-2mm}
\begin{equation}
    \widetilde{f}_k(u,v) = \int_{-\infty}^{+\infty} \int_{-\infty}^{+\infty} \widetilde{f}_{k}e^{-i(u\widetilde{f}^{i}_k + v\widetilde{f}^{j}_k)} d\widetilde{f}^{i}_k d\widetilde{f}^{j}_k,
\vspace{-2mm}
\end{equation}
where $u$ and $v$ denote the spatial frequencies of the Fourier transform. Due to only a small number of cells being obscured by interference, a high-pass filter is applied to the frequency domain map for the removal of these low-frequency information:
\vspace{-2mm}
\begin{equation}
    f^{'}_k = \xi^{-1}[H_{high}(u,v)\widetilde{f}_k(u,v)],
\vspace{-2mm}
\end{equation}
where $\xi^{-1}$ denotes the Inverse Discrete Fourier Transform (IDFT), $f^{'}_k$ denotes the output feature map of FTB after interference information got eliminated and $H_{high}(u,v)$ denotes the high-pass filter.
$f^{'}_k$ is then re-sent to the attention block (AB) to get the final feature embedding, which eventually is passed to a prediction layer to get the final prediction mask $y^{t}_k (k \in {3, 4, 5})$.

\subsection{Optimization}\label{sec3.2.3}
For optimization, we set binary cross entropy (BCE) loss as the target function for both coarse result maps $y^c_k (k \in {3, 4, 5})$ and final prediction masks $y^{t}_k (k \in {3, 4, 5})$. Thus, the overall loss function for the training process could be written as 
\vspace{-2mm}
\begin{equation}
    \mathcal{L} = \sum^{5}_{k=3} \mathcal{L}_{BCE}(y^c_k, y) + \sum^{5}_{k=3} \mathcal{L}_{BCE}(y^t_k, y),
\vspace{-2mm}
\end{equation}
where $y$ denotes the ground-truth mask and $L_{BCE}$ denotes the binary cross entropy loss.

\section{Experiments and results}\label{sec4}

\begin{figure*}[!t]
\centerline{\includegraphics[width=1.0\textwidth]{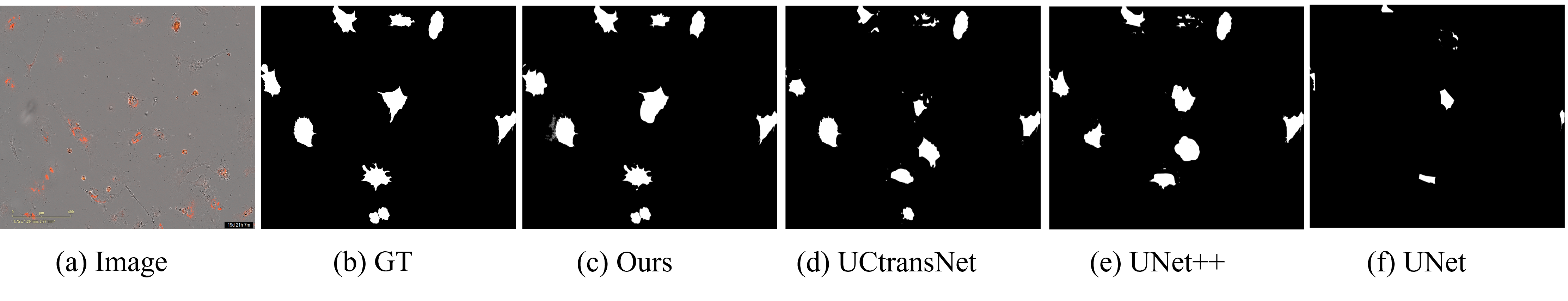}}\vspace{-4.5mm}
\caption{Visualization of astrocyte segmentation results with (a) as the input of the network. (b) is a ground-truth (GT) segmentation mask and (c-f) represent the segmentation masks generated by our proposed network, UCtransNet, UNet++ and UNet, respectively.}\vspace{-0.5cm}

\label{visualization}
\end{figure*}

\subsection{Experimental Settings}\label{sec4.1}
\xhdr{Implementation Details}We use the ResNet-50 pre-trained on ImageNet as our backbone. The whole network is trained on a single NVIDIA A100 Tensor Core GPU with 80GB memory. For training, all images are resized to $1024 \times 1024$ for computational efficiency. We choose the Adam optimizer with an initial learning rate set to 0.001 for optimization. The batch size is 8 and the epoch number is 400. The learning rate is decayed by $50\%$ every 100 epochs.

\xhdr{Evaluation Metrics}We adopt two commonly used metrics, \ie, mean of intersection over union (mIoU) and dice coefficient (Dice), as the evaluation metrics.

\xhdr{Baselines}We compare our proposed model with seven state-of-the-art segmentation approaches based on the following criteria: a) classical framework for image segmentation, b) recently published, and c) achieve SOTA performance for medical image segmentation.

\begin{table}[!t]
\centering
\caption{Quantitative comparison segmentation performance on IAI704. $\uparrow$ represents that the higher is better. }
\renewcommand{\arraystretch}{1.15}
\resizebox{0.7\linewidth}{!}{
\begin{tabular}{l|c|c}
\hline
Method& mIoU $\uparrow$ & Dice $\uparrow$ \\
\cline{1-3}
UNet~\pub{MICCAI2015}~\cite{ronneberger2015u}& 67.1& 74.3  \\
\hline
Deeplab~\pub{ECCV2018}~\cite{chen2018encoder} & 58.4& 63.5\\
\hline
UNet++~\pub{DLMIA18}~\cite{zhou2018unet++}& 60.2& 68.4  \\
\hline
TransUnet~\pub{arXiv2021}~\cite{chen2021transunet}& 56.8& 61.0  \\
\hline
PraNet~\pub{MICCAI2020}~\cite{fan2020pranet}& 70.6& 77.9  \\
\hline
SwinUnet~\pub{ECCV2022}~\cite{cao2022swin}& 53.7& 59.4  \\
\hline
UCtransNet~\pub{AAAI2022}~\cite{wang2022uctransnet}& 75.7& 82.1  \\
\hline
\textbf{Ours}& \textbf{80.8}& \textbf{86.2}  \\
\hline
\end{tabular}
}\vspace{-0.5cm}

\label{tab1}

\end{table}

\subsection{Comparisons with State-of-the-arts}\label{sec4.2}
We report the results of our FDNet on the IAI704 dataset in Table~\ref{tab1}, from which it can be observed that the proposed network outperforms all the baseline models to a large extent. Due to the fact that ``Deeplab'' and ``Unet'' are designed for regular images, these two classic segmentation methods don't perform well on our proposed dataset, in which the target and background are highly similar. Compared to ``UCtransNet'', our proposed FDNet performs much better on the astrocyte segmentation, which leads to substantial performance gains in $mIoU$ (\ie, $75.7 \rightarrow 80.8$) and $Dice$ (\ie, $82.1 \rightarrow 86.2$). Fig.~\ref{visualization} shows the segmentation results of the astrocyte segmentation on IAI704.

\subsection{Ablation study}\label{sec4.3}

We set a bunch of ablation studies to evaluate the effectiveness of different modules in the proposed FDNet. We use No.1 $\sim$ No.4 to represent different experimental settings. As can be seen from Table~\ref{tab2}, using backbone only (No.1) will lead to a bad performance on astrocyte segmentation. Compared No.2 with No.1, the segmentation performance boost due to the CIF module fuses multi-scale features which can expand the receptive field of the network. By making the model pay more attention to key areas or eliminating interfences, No.3 and No.4 have both improved the segmentation performance (\ie, $mIoU$ increased by $17\%$ for No.3 and $Dice$ improved from $61.3 \rightarrow 74.1$ for No.4). Although adding different modules separately could boost the astrocyte segmentation performance, our proposed FDNet achieves the best performance with combining all these modules together. Noted that to further prove the effectiveness of our proposed FTB module, we set a separate experimental setting by adding the FTB module before the last convolution layer of the ``UCtransNet''. It can be seen from the last row of Table~\ref{tab2} that although there is a significant improvement on both $mIoU$ (\ie, $75.7 \rightarrow 78.6$) and $Dice$ (\ie, $82.1 \rightarrow 83.4$) compared with its original framework, our FDNet still performs better. These results demonstrate the effectiveness of CIF, AB, and FTB in our proposed framework for astrocyte segmentation.

\begin{table}[t!]
	\centering
\caption{ Ablation study of the key components on proposed dataset IAI704 {\tt test}. No.1 $\sim$ No.4 are used to denote different combination of the key components. }
\renewcommand{\arraystretch}{1.15}
\resizebox{0.9\linewidth}{!}{

		\begin{tabular}{l|cccc|cc}
			\hline
			\multicolumn{1}{l}{}&\multicolumn{4}{c|}{Candidate}
			& \multicolumn{2}{c}{IAI704-{\tt test}} \\
			\cline{1-7}
			\multicolumn{1}{l}{}&ResNet &CIF &AB &FTB &$mIoU\uparrow$ &Dice$\uparrow$ \\
			\hline
			\multicolumn{1}{l|}{No.1}&\ding{52}&&&&50.4&54.7\\
			\multicolumn{1}{l|}{No.2}&\ding{52}&\ding{52}&&&58.1&61.3\\
			\multicolumn{1}{l|}{No.3}&\ding{52}&\ding{52}&\ding{52}&&68.5&73.8\\
			\multicolumn{1}{l|}{No.4}&\ding{52}&\ding{52}&&\ding{52}&69.7&74.1\\
                \multicolumn{1}{l|}{\textbf{Ours}}&\ding{52}&\ding{52}&\ding{52}&\ding{52} &\textbf{80.8}&\textbf{86.2}\\
			\hline
                \multicolumn{1}{l|}{UCtransNet}&&&&\ding{52} &78.6&83.4\\
                \hline
		\end{tabular}
  }\vspace{-0.5cm}
\label{tab2}
\end{table}

\section{Conclusion}\label{sec5}
In this paper, we introduced a new task named astrocyte segmentation. This task aims to segment astrocytes from Incucyte images at their different differentiation stages. To handle this task, we provided a new annotated IAI704 dataset with a frequency domain denoising network named FDNet. Specifically, our FDNet consists of an attention block module for accurate localization and a Fourier Transform block module for interference information removal. 
Extensive experiments prove that our method can successfully segment astrocytes in Incucyte images. Furthermore, we expect our paper could open up the possibility of doing deeper Incucyte image analysis for other cell types in the future and shedding insights for neurodegenerative diseases.

\section{Acknowledgments}
\label{sec:acknowledgments}

The authors acknowledge Cedars-Sinai Medical Center’s David and Janet Polak Foundation Stem Cell Core Laboratory for providing the HD iPSC line, as well as Professor Lesley Jones and Dr Thomas Massey from Cardiff University, UK for providing the isogenic control HD iPSC line. The lentiviral vectors were received from Professor Lezanne Ooi at University of Wollongong, Australia. The authors also acknowledge NVIDIA and its research support team for the help provided to conduct this work. We'd like to acknowledge the generosity and support of our silent donor that makes this research into Huntington's disease possible. This study was also supported by the Medical Advances Without Animals (MAWA) Trust. The aim of MAWA is to advance medical science and improve human health and therapeutic interventions without the use of animals or animal products.This work was also partially supported by the Australian Research Council (ARC) Industrial Transformation Training Centres (IITC) for Innovative Composites for the Future of Sustainable Mining Equipment under Grant IC220100028.

\bibliographystyle{IEEEbib}
\bibliography{strings,refs}

\end{document}